\documentclass[conference]{IEEEtran}
\IEEEoverridecommandlockouts
\usepackage{cite}
\usepackage{amsmath,amssymb,amsfonts}
\usepackage{algorithmic}
\usepackage{graphicx}
\usepackage{textcomp}
\usepackage{xcolor}

\usepackage{hyperref}       
\usepackage{url}            
\usepackage{subcaption}

\def\BibTeX{{\rm B\kern-.05em{\sc i\kern-.025em b}\kern-.08em
    T\kern-.1667em\lower.7ex\hbox{E}\kern-.125emX}}
\begin{document}

\title{Quantifying Spatial Domain Explanations in BCI using Earth Mover's Distance\\

}


\author{\IEEEauthorblockN{Param Rajpura}
\IEEEauthorblockA{{Human-AI Interaction (HAIx) Lab} \\
\textit{IIT Gandhinagar, India}\\
rajpuraparam@iitgn.ac.in}
\and
\IEEEauthorblockN{Hubert Cecotti}
\IEEEauthorblockA{{Department of Computer Science} \\
\textit{California State University,Fresno, USA}\\
hcecotti@csufresno.edu}
\and
\IEEEauthorblockN{Yogesh Kumar Meena}
\IEEEauthorblockA{{Human-AI Interaction (HAIx) Lab} \\
\textit{IIT Gandhinagar, India}\\
yk.meena@iitgn.ac.in}
}

\maketitle

\begin{abstract}


Brain-computer interface (BCI) systems facilitate unique communication between humans and computers, benefiting severely disabled individuals. Despite decades of research, BCIs are not fully integrated into clinical and commercial settings. It's crucial to assess and explain BCI performance, offering clear explanations for potential users to avoid frustration when it doesn't work as expected. This work investigates the efficacy of different deep learning and Riemannian geometry-based classification models in the context of motor imagery (MI) based BCI using electroencephalography (EEG). We then propose an optimal transport theory-based approach using earth mover's distance (EMD) to quantify the comparison of the feature relevance map with the domain knowledge of neuroscience. For this, we utilized explainable AI (XAI) techniques for generating feature relevance in the spatial domain to identify important channels for model outcomes. Three state-of-the-art models are implemented - 1) Riemannian geometry-based classifier, 2) EEGNet, and 3) EEG Conformer, and the observed trend in the model's accuracy across different architectures on the dataset correlates with the proposed feature relevance metrics. The models with diverse architectures perform significantly better when trained on channels relevant to motor imagery than data-driven channel selection. This work focuses attention on the necessity for interpretability and incorporating metrics beyond accuracy, underscores the value of combining domain knowledge and quantifying model interpretations with data-driven approaches in creating reliable and robust Brain-Computer Interfaces (BCIs).




\end{abstract}


\section{Introduction}

A Brain-Computer Interface (BCI), also known as a Brain-Machine Interface (BMI) or Neural Interface, is a technology that establishes a direct communication pathway between the brain and an external device, such as a computer or prosthetic limb~\cite{rashid2020current}. The primary goal of a BCI is to enable the exchange of information between the brain and an external system, bypassing traditional sensory or motor pathways like muscles or nerves.

BCIs can be designed for various purposes. BCIs can assist individuals with disabilities by allowing them to control devices or communicate through direct brain signals. This is particularly relevant for people with paralysis or severe motor impairments~\cite{belkacem2020brain}. BCIs can be used to provide individuals with real-time feedback on their brain activity. This can be employed in areas like Neurotherapy, where individuals can learn to regulate their brain activity for therapeutic purposes~\cite{cantillo2021brain}. BCIs can also enhance virtual or augmented reality experiences by allowing users to interact with digital environments using their thoughts.
BCIs are valuable tools in neuroscience research. They can be used to study brain function, cognitive processes, and neurological disorders~\cite{mudgal2020brain}. In the medical field, BCIs may be employed for diagnostic purposes or as part of rehabilitation strategies~\cite{zhuang2020state}. In such a case, it is necessary to have results that can be fully explainable.


Selecting the right sensors for a BCI is important for achieving the desired performance, user acceptance, and practicality in various applications\cite{alotaiby2015review}. In BCIs designed for practical applications, user experience is critical. The sensors should be non-intrusive, easy to wear, and comfortable for the user. This is particularly important for applications that require long-term use, such as assistive technologies or rehabilitation devices. The spatial resolution of sensors determines how finely the brain activity can be localized. Some sensors may have better spatial resolution, allowing for more precise identification of the brain regions involved in a particular task or thought\cite{lal2004support}. This is especially important for applications that require detailed mapping of neural activity, such as neuroscientific research or precise control in neuroprosthetics.

The recent approaches towards adopting deep learning (DL) based Motor Imagery (MI) models in BCI research are driven by the advantages associated with learning complex tasks in an end-to-end fashion \cite{schirrmeister2017deep,lawhern2018eegnet}. This paradigm shift accompanies a challenge in offering interpretable and robust solutions\cite{tjoa2020survey,doshi2017towards,tonekaboni2019clinicians,ribeiro2016should}. Explainable AI(XAI) \cite{gunning2017explainable} has been applied to BCIs to improve performance 
\cite{song2022eeg,salami2022eeg,lawhern2018eegnet} or for optimal channel selection \cite{nagarajan2022relevance}. Though XAI approaches are utilized to justify the model performance, the explanations from a specific model architecture are subjectively interpreted in isolation. Evaluating model explanations with quantified metrics is a stepping stone toward interpretable models. Ravindran and Contreras-Vidal \cite{sujatha2023empirical} used a simulation framework to evaluate the robustness and sensitivity of twelve back-propagation-based visualization methods by comparing them to ground truth features of the simulated EEG data. Cui et al. \cite{cui2023towards} quantitatively evaluate seven different deep
interpretation techniques across different models and datasets for EEG-based BCI. Both works evaluate the robustness of XAI approaches by perturbing the data. Complementing these insights, we compare and ground these insights with the domain knowledge in neuroscience.   

The key contributions of the paper are: 1) a novel approach using Earth Mover's distance metric based on optimal transport theory is proposed to analyze and compare the spatial domain explanations for predicted outcomes of the models used in EEG-based BCI with existing domain knowledge; 2) a detailed benchmark of state-of-the-art models is provided on a dataset with 109 participants to evaluate their performance and explanations.
The rest of the paper is organized as follows. The dataset and the different classification algorithms are presented in Section~\ref{sec:methods}. The classification performance is given in Section~\ref{sec:results}. The impact of the results on eXplainable AI for BCI is discussed in Section~\ref{sec:discussion}. The key findings are summarised in Section~\ref{sec:conclusion}.

\section{Methods}
\label{sec:methods}

\subsection{Dataset Description}

The EEGMMID-Physionet dataset is used in this work\cite{goldberger2000physiobank,schalk2004bci2000}. It is one of the largest datasets on motor imagery (MI) and movement, with 64-channel EEG recordings collected from 109 participants.  We considered the two tasks: 1) open and close left or right fist and 2) imagine opening and closing left or right fist. It is a binary classification problem, i.e., with two classes.

We merged both tasks to augment the dataset. Each subject, on average, did 93 trials, each 4 seconds long. We divided each 4-second trial into four 1-second length runs, resulting in 160 sample recordings in each sequence and an average of 370 sequences for each subject. A bandpass filter of 8-30~Hz was applied to remove any noise in the EEG signals, including the mu (8-12~Hz) and beta (12-30~Hz) bands that contain event synchronization/desynchronization (ERS/ERD) information~\cite{meena2015towards, o2014exploring, chowdhury2018active}. Each sequence was labeled based on the activity for left and right fist movement/imagery respectively. 

\subsection{Model Architecture and Training}

Three diverse model architectures are explored and utilised to evaluate and compare explanations from different families of classification methods. 1) Riemannian geometry-based classifier (density-based classifier using the distance to the mean)\cite{barachant2011multiclass}, 2) Convolutional feedforward neural network architecture: EEGNet (discriminant approach)~\cite{lawhern2018eegnet}, and 3) Transformer and self-attention-inspired architecture: EEG Conformer (discriminant approach)~\cite{song2022eeg}.

\subsubsection{Minimum distance to mean (MDM) using Riemannian distance}

The proposed methodology by Barachant et al. \cite{barachant2011multiclass} involves utilizing covariance matrices derived from raw EEG signals. Each value in these matrices represents the variance across two EEG channels. Since these matrices are symmetric positive definite (SPD), the underlying data structure is non-euclidean. Leveraging Riemannian geometry approaches facilitates directly classifying EEG signals without spatial filtering. It is widely successful in classifying relevant signals for motor imagery using EEG \cite{yger2016riemannian,congedo2017riemannian}.

To derive an explanation using feature relevance, we extract the most relevant channels for predictions, as proposed by Barachant and Bonnet \cite{barachant2011channel} and implemented in PyRiemannian library~\cite{pyriemann}, for each class, a centroid is estimated, i.e., the average across examples of a given class, and the channel selection is based on the maximization of the distance between centroids. The principle is related to the k-nearest neighbor but instead of estimating the distance to examples, the distance is estimated to the centroids. Backward elimination removes the electrode that carries the least distance from the subset at each iteration. The process is repeated until the required number of electrode channels remains in the subset.

\subsubsection{EEGNet}

EEGNet is a convolutional neural network consisting of two convolutional steps in subsequent layers~\cite{lawhern2018eegnet}. The first layer with temporal convolution learns frequency filters, followed by depthwise convolution to learn spatial filters for specific frequencies. The architecture then uses a depthwise convolution and pointwise convolution to classify the features using a softmax layer. We follow the experimental setup based on its Torch library \cite{paszke2019pytorch} implementation in the TorchEEG package.

\subsubsection{EEG Conformer}

EEG Conformer has three stages: the convolutional stage consisting of two one-dimensional convolutions to learn temporal and spatial dependencies, followed by the self-attention module to learn global dependencies, and finally, the classifier stage, two fully connected layers along with the softmax function to predict the activity in hands or limb movement/imagery. We follow the experimental setup and implementation of Song et al. \cite{song2022eeg} for this work. The model is trained separately for each participant with the same hyperparameters except for the second convolutional stage, where the kernel size is adjusted as per the number of EEG channels used in the dataset. 

\subsection{Explainable AI (XAI) technique: GradCAM}

For the visualization and explanation interface, leveraging the statistical data-driven observations, we use Gradient-weighted Class Activation Mapping (Grad-CAM) \cite{selvaraju2017grad} to generate the feature relevance maps (montage images highlighting significant EEG channels based on feature relevance scores from GradCAM).

GradCAM is a class-specific approach combining the gradients and the input features to interpret the results. Such a class discriminative approach enables the comparison between the features learned by the model for the two tasks, respectively. 

For EEGNet architecture, we use the gradient information from the first convolution layer that represents the feature relevance for each channel at different frequencies or temporal spans. For EEG Conformer, we use the gradients from the self-attention module as implemented in the original work\cite{song2022eeg,selvaraju2017grad}. For both models, to generate the explanations using the relevance scores, the required number of the most relevant channels is extracted.


\subsection{Experimental Setup}

For our first experiment, three models, MDM, EEGConformer, and EEGNet, are trained using all 64-channel data for each subject. The trained models are evaluated on the randomly shuffled epochs selected for the test set with a fixed seed value. Considering the MDM model as a baseline, we choose 14 participants among the 109 in the dataset with task accuracy at least 10\% higher than chance level accuracy. This step ensures that the subsequent analysis is unbiased and robust. Further, to evaluate the feature relevance among the 64 channels, we use Backward elimination for the MDM classifier and GradCAM technique for EEGConformer and EEGNet. The most relevant 21 channels, as per feature relevance scores for left and right-hand movement/imagery, are used to train the models. Finally, to compare feature relevance with domain knowledge, the models are trained considering the 21 channels positioned near the motor cortical regions, especially central, frontal-central, and parietal regions \cite{mcfarland2000mu}. These combinations result in three configurations: $1)$ Using all 64 EEG channels, $2)$ Using 21 feature-relevant channels, and $3)$ Using 21 motor imagery (MI) relevant channels for each classification model respectively, totalling nine unique configurations of classification models.

\begin{table*}
\centering
\resizebox{\textwidth}{!}{
\begin{tabular}{c | c | c c c | c c c | c c c}
\textbf{ID} & \multicolumn{1}{c|}{\textbf{\begin{tabular}[c]{@{}c@{}}Chance level\\ Accuracy\end{tabular}}} & \multicolumn{3}{c|}{\textbf{Accuracy using all EEG channels}} & \multicolumn{3}{c|}{\textbf{\begin{tabular}[c]{@{}c@{}}Accuracy using 21 \\ MI relevant channels\end{tabular} }} & \multicolumn{3}{c}{\textbf{\begin{tabular}[c]{@{}c@{}}Accuracy using top 21 \\  feature relevant channels\end{tabular} }} \\
\hline
 &  & \multicolumn{1}{c}{\textbf{\begin{tabular}[c]{@{}c@{}}Overall\end{tabular}}} & \multicolumn{1}{c}{\textbf{\begin{tabular}[c]{@{}c@{}}Left fist\end{tabular}}} & \multicolumn{1}{c}{\textbf{\begin{tabular}[c]{@{}c@{}}Right fist\end{tabular}}} & \multicolumn{1}{|c}{\textbf{\begin{tabular}[c]{@{}c@{}}Overall\end{tabular}}} & \multicolumn{1}{c}{\textbf{\begin{tabular}[c]{@{}c@{}}Left fist\end{tabular}}} & \multicolumn{1}{c|}{\textbf{\begin{tabular}[c]{@{}c@{}}Right fist\end{tabular}}} & \multicolumn{1}{c}{\textbf{\begin{tabular}[c]{@{}c@{}}Overall\end{tabular}}} & \multicolumn{1}{c}{\textbf{\begin{tabular}[c]{@{}c@{}}Left fist\end{tabular}}} & \multicolumn{1}{c}{\textbf{\begin{tabular}[c]{@{}c@{}}Right fist\end{tabular}}} \\
7 & 59.86 & 83.87 & 83.93 & 83.78 & 75.27 & 71.43 & 81.08 & 74.19 & 75.00 & 72.97 \\
12 & 57.61 & 73.12 & 77.78 & 66.67 & 69.89 & 74.07 & 64.10 & 73.12 & 74.07 & 71.79 \\
22 & 58.78 & 73.12 & 80.00 & 63.16 & 61.29 & 63.64 & 57.89 & 63.44 & 65.45 & 60.53 \\
42 & 56.88 & 78.49 & 83.02 & 72.50 & 75.27 & 79.25 & 70.00 & 77.42 & 75.47 & 80.00 \\
43 & 57.61 & 68.82 & 64.81 & 74.36 & 64.52 & 59.26 & 71.79 & 61.29 & 57.41 & 66.67 \\
48 & 56.52 & 77.42 & 81.13 & 72.50 & 75.27 & 79.25 & 70.00 & 72.04 & 75.47 & 67.50 \\
49 & 58.70 & 72.04 & 81.48 & 58.97 & 73.12 & 74.07 & 71.79 & 65.59 & 81.48 & 43.59 \\
53 & 57.25 & 74.19 & 77.36 & 70.00 & 62.37 & 58.49 & 67.50 & 73.12 & 69.81 & 77.50 \\
70 & 59.06 & 69.89 & 72.73 & 65.79 & 60.22 & 52.73 & 71.05 & 67.74 & 65.45 & 71.05 \\
80 & 59.06 & 70.97 & 78.18 & 60.53 & 60.22 & 65.45 & 52.63 & 64.52 & 63.64 & 65.79 \\
82 & 57.97 & 70.97 & 85.19 & 51.28 & 66.67 & 77.78 & 51.28 & 67.74 & 79.63 & 51.28 \\
85 & 58.70 & 70.97 & 72.22 & 69.23 & 74.19 & 75.93 & 71.79 & 65.59 & 62.96 & 69.23 \\
94 & 58.33 & 77.42 & 79.63 & 74.36 & 84.95 & 90.74 & 76.92 & 75.27 & 81.48 & 66.67 \\
102 & 57.61 & 69.57 & 71.70 & 66.67 & 71.74 & 73.58 & 69.23 & 58.70 & 56.60 & 61.54 \\
\hline 
\textbf{Mean$\pm$SD} & \textbf{58.14$\pm$0.94} & \textbf{73.63$\pm$4.26} & 77.80$\pm$5.65 & 67.84$\pm$7.96 & \textbf{69.64$\pm$7.35} & 71.12$\pm$10.13 & 67.65$\pm$8.52 & \textbf{68.56$\pm$5.69} & 70.28$\pm$8.44 & 66.15$\pm$9.68 \\
\hline \\
\end{tabular}
}
\caption{Chance level accuracy compared to the performance of MDM classifier using Riemannian distance on covariance matrix using 1) all 64 EEG channels, 2) using 21 Motor Imagery and movement-related EEG channels leveraging the domain knowledge and 3) using 21 most relevant EEG channels identified by relevance scores}
\label{tab:riem_accTable}
\end{table*}

\subsection{Performance evaluation}

Due to the unbalanced prior probability of the classes, we define the accuracy for each class as the number of correctly detected examples for a given class by the total number of examples of this class. The overall accuracy is weighted based on the prior class probabilities, calculated using the average of recall obtained on each class to avoid any bias in an imbalanced dataset \cite{brodersen2010balanced}. All the results are given as percentages.

\subsection{Problem Formulation for Quantifying explanations}

To quantify the comparison of relevant channels when the three models are trained on all 64 EEG channels, the montage (as visualized in \autoref{fig:montage_comparison}) as per the international 10-10 system \cite{schalk2004bci2000} is projected onto a matrix of order $N$x$N$ where $N$ is the maximum number of electrodes in any of the two axes on the projected space. For this case, $N=11$ corresponds to eleven electrodes corresponding to the central and temporal regions. Hence, each element in the matrix represents a spatial location that may correspond to the position of an electrode in the projected space.
For simplicity, the channels considered in the top 21 features extracted using the XAI technique are marked with 1, while the rest of the locations are marked as 0, denoting the feature relevance as a binary value. Such a matrix (as visualized in \autoref{fig:spatial_map}) can now be used to compare the explanations for a model outcome that are derived in terms of the feature relevance of an EEG channel.

\subsection{Proposed Approach for Quantifying explanations}
We compare such matrices by calculating the Earth Mover's distance \cite{bonneel2011displacement} implemented using Python Optimal Transport library \cite{flamary2021pot}. This method involves assessing the dissimilarity between two matrices, measuring the minimum work required to transform one into the other. Earth Mover's distance(EMD), also known as Wasserstein distance, evaluates the optimal mass transportation plan, considering both values and spatial arrangements. This approach provides a robust metric for matrix dissimilarity, capturing both local and global differences.

Posing the dissimilarity as an optimal transport problem, let \( P \) and \( Q \) represent two spatial maps, where each map is a set of spatial elements, i.e., the relevance of each EEG channel location, denoted as \( p_i \) and \( q_j \). Here, \( n \) and \( m \) are the number of channels in each map.

The ground distance matrix \( C \) signifies the cost of moving mass from one channel location in \( P \) to another in \( Q \). \( C_{ij} \) represents the cost of transporting one unit of mass from \( p_i \) to \( q_j \), capturing the pairwise distances between channel locations.

The transportation matrix \( T \) outlines the optimal plan for moving mass from \( P \) to \( Q \). Each \( T_{ij} \) element indicates the amount of mass transported from \( p_i \) to \( q_j \). The objective is to find \( T \) that minimizes the total transportation cost. 

The EMD is then calculated as the sum of the element-wise product of \( C \) and \( T \):

\[ \text{EMD}(P, Q) = \sum_{i=1}^{n} \sum_{j=1}^{m} C_{ij} \cdot T_{ij} \]

Subject to the constraints:

1. The sum of mass transported from any \( p_i \) to all \( q_j \) must be equal to \( p_i \): \( \sum_{j=1}^{m} T_{ij} = p_i \) for all \( i \).

2. The sum of mass transported to any \( q_j \) from all \( p_i \) must be equal to \( q_j \): \( \sum_{i=1}^{n} T_{ij} = q_j \) for all \( j \).

\begin{table*}
\centering
\resizebox{\textwidth}{!}{
\begin{tabular}{c | c | c c c | c c c | c c c}
\textbf{ID} & \multicolumn{1}{c|}{\textbf{\begin{tabular}[c]{@{}c@{}}Chance level\\ Accuracy\end{tabular}}} & \multicolumn{3}{c|}{\textbf{Accuracy using all EEG channels}} & \multicolumn{3}{c|}{\textbf{\begin{tabular}[c]{@{}c@{}}Accuracy using 21 \\ MI relevant channels\end{tabular} }} & \multicolumn{3}{c}{\textbf{\begin{tabular}[c]{@{}c@{}}Accuracy using top 21 relevant \\ channels from GradCAM\end{tabular} }} \\
\hline
 &  & \multicolumn{1}{c}{\textbf{\begin{tabular}[c]{@{}c@{}}Overall\end{tabular}}} & \multicolumn{1}{c}{\textbf{\begin{tabular}[c]{@{}c@{}}Left fist\end{tabular}}} & \multicolumn{1}{c}{\textbf{\begin{tabular}[c]{@{}c@{}}Right fist\end{tabular}}} & \multicolumn{1}{|c}{\textbf{\begin{tabular}[c]{@{}c@{}}Overall\end{tabular}}} & \multicolumn{1}{c}{\textbf{\begin{tabular}[c]{@{}c@{}}Left fist\end{tabular}}} & \multicolumn{1}{c|}{\textbf{\begin{tabular}[c]{@{}c@{}}Right fist\end{tabular}}} & \multicolumn{1}{c}{\textbf{\begin{tabular}[c]{@{}c@{}}Overall\end{tabular}}} & \multicolumn{1}{c}{\textbf{\begin{tabular}[c]{@{}c@{}}Left fist\end{tabular}}} & \multicolumn{1}{c}{\textbf{\begin{tabular}[c]{@{}c@{}}Right fist\end{tabular}}} \\
7 & 59.86 & 73.12 & 69.64 & 78.38 & 80.65 & 78.57 & 83.78 & 60.22 & 76.79 & 35.14 \\
12 & 57.61 & 74.19 & 75.93 & 71.79 & 67.74 & 70.37 & 64.10 & 75.27 & 81.48 & 66.67 \\
22 & 58.78 & 68.82 & 67.27 & 71.05 & 60.22 & 65.45 & 52.63 & 70.97 & 85.45 & 50.00 \\
42 & 56.88 & 78.49 & 83.02 & 72.50 & 82.80 & 84.91 & 80.00 & 67.74 & 66.04 & 70.00 \\
43 & 57.61 & 62.37 & 72.22 & 48.72 & 62.37 & 72.22 & 48.72 & 63.44 & 87.04 & 30.77 \\
48 & 56.52 & 74.19 & 81.13 & 65.00 & 73.12 & 69.81 & 77.50 & 67.74 & 75.47 & 57.50 \\
49 & 58.70 & 69.89 & 81.48 & 53.85 & 74.19 & 77.78 & 69.23 & 73.12 & 75.93 & 69.23 \\
53 & 57.25 & 69.89 & 77.36 & 60.00 & 60.22 & 67.92 & 50.00 & 75.27 & 84.91 & 62.50 \\
70 & 59.06 & 61.29 & 85.45 & 26.32 & 56.99 & 58.18 & 55.26 & 59.14 & 78.18 & 31.58 \\
80 & 59.06 & 60.22 & 70.91 & 44.74 & 59.14 & 61.82 & 55.26 & 56.99 & 69.09 & 39.47 \\
82 & 57.97 & 53.76 & 62.96 & 41.03 & 59.14 & 64.81 & 51.28 & 60.22 & 57.41 & 64.10 \\
85 & 58.70 & 68.82 & 70.37 & 66.67 & 58.06 & 57.41 & 58.97 & 55.91 & 61.11 & 48.72 \\
94 & 58.33 & 79.57 & 92.59 & 61.54 & 77.42 & 85.19 & 66.67 & 68.82 & 79.63 & 53.85 \\
102 & 57.61 & 66.30 & 58.49 & 76.92 & 56.52 & 77.36 & 28.21 & 58.70 & 54.72 & 64.10 \\
\hline 
\textbf{Mean$\pm$SD} & \textbf{58.14$\pm$0.94} & \textbf{68.64$\pm$7.30} & 74.92$\pm$9.29 & 59.89$\pm$15.21 & \textbf{66.33$\pm$9.42} & 70.84$\pm$9.00 & 60.12$\pm$14.76 & \textbf{65.25$\pm$6.85} & 73.80$\pm$10.52 & 53.12$\pm$14.09 \\
\hline \\
\end{tabular}
}
\caption{Chance level accuracy compared to the performance of EEG Conformer model using 1) all 64 EEG channels, 2) using 21 Motor Imagery and movement-related EEG channels leveraging the domain knowledge, and 3) using 21 most relevant EEG channels identified by GradCAM.}
\label{tab:conf_accTable}
\end{table*}

\section{Results}
\label{sec:results} 

A comprehensive overview of model performance under different setups, using all 64 EEG channels, 21 feature-relevant channels, and 21 motor imagery (MI) relevant channels, respectively, for the three model architectures is represented in \autoref{tab:riem_accTable}, \autoref{tab:conf_accTable}, and \autoref{tab:eegnet_accTable}, respectively. MDM classifier using Riemannian distance achieves 73.63\% accuracy using all channels. When trained on 21 MI-relevant channels, the performance reduces by 3.99\% to 69.64\%  (\textit{p=0.0279} using Wilcoxon signed-rank test). Using the backward elimination method for channel selection, accuracy is 68.56\%, decreasing by 5.07\% (\textit{p=0.0014} using Wilcoxon signed-rank test).
A similar trend is observed for EEGConformer and EEGNet across the configurations; however, comparing the three model's performance MDM classifier outperforms EEGConformer by 4.99\% (\textit{p=0.0029} using Wilcoxon signed-rank test) and EEGNet by 6.61\% (\textit{p=0.0028} using Wilcoxon signed-rank test) using all channels.

\begin{table*}
\centering
\resizebox{\textwidth}{!}{
\begin{tabular}{c | c | c c c | c c c | c c c}
\textbf{ID} & \multicolumn{1}{c|}{\textbf{\begin{tabular}[c]{@{}c@{}}Chance level\\ Accuracy\end{tabular}}} & \multicolumn{3}{c|}{\textbf{Accuracy using all EEG channels}} & \multicolumn{3}{c|}{\textbf{\begin{tabular}[c]{@{}c@{}}Accuracy using 21 \\ MI relevant channels\end{tabular} }} & \multicolumn{3}{c}{\textbf{\begin{tabular}[c]{@{}c@{}}Accuracy using top 21 relevant \\ channels from GradCAM\end{tabular} }} \\
\hline
 &  & \multicolumn{1}{c}{\textbf{\begin{tabular}[c]{@{}c@{}}Overall\end{tabular}}} & \multicolumn{1}{c}{\textbf{\begin{tabular}[c]{@{}c@{}}Left fist\end{tabular}}} & \multicolumn{1}{c}{\textbf{\begin{tabular}[c]{@{}c@{}}Right fist\end{tabular}}} & \multicolumn{1}{|c}{\textbf{\begin{tabular}[c]{@{}c@{}}Overall\end{tabular}}} & \multicolumn{1}{c}{\textbf{\begin{tabular}[c]{@{}c@{}}Left fist\end{tabular}}} & \multicolumn{1}{c|}{\textbf{\begin{tabular}[c]{@{}c@{}}Right fist\end{tabular}}} & \multicolumn{1}{c}{\textbf{\begin{tabular}[c]{@{}c@{}}Overall\end{tabular}}} & \multicolumn{1}{c}{\textbf{\begin{tabular}[c]{@{}c@{}}Left fist\end{tabular}}} & \multicolumn{1}{c}{\textbf{\begin{tabular}[c]{@{}c@{}}Right fist\end{tabular}}} \\
7 & 59.86 & 78.49 & 83.93 & 70.27 & 81.72 & 82.14 & 81.08 & 66.67 & 75.00 & 54.05 \\
12 & 57.61 & 65.59 & 61.11 & 71.79 & 58.06 & 53.70 & 64.10 & 62.37 & 79.63 & 38.46 \\
22 & 58.78 & 59.14 & 63.64 & 52.63 & 61.29 & 58.18 & 65.79 & 56.99 & 65.45 & 44.74 \\
42 & 56.88 & 68.82 & 69.81 & 67.50 & 78.49 & 79.25 & 77.50 & 56.99 & 66.04 & 45.00 \\
43 & 57.61 & 63.44 & 64.81 & 61.54 & 58.06 & 53.70 & 64.10 & 59.14 & 62.96 & 53.85 \\
48 & 56.52 & 70.97 & 75.47 & 65.00 & 63.44 & 62.26 & 65.00 & 64.52 & 73.58 & 52.50 \\
49 & 58.70 & 75.27 & 77.78 & 71.79 & 73.12 & 66.67 & 82.05 & 73.12 & 81.48 & 61.54 \\
53 & 57.25 & 68.82 & 71.70 & 65.00 & 64.52 & 66.04 & 62.50 & 65.59 & 73.58 & 55.00 \\
70 & 59.06 & 59.14 & 56.36 & 63.16 & 58.06 & 63.64 & 50.00 & 61.29 & 63.64 & 57.89 \\
80 & 59.06 & 60.22 & 63.64 & 55.26 & 59.14 & 60.00 & 57.89 & 51.61 & 54.55 & 47.37 \\
82 & 57.97 & 62.37 & 72.22 & 48.72 & 54.84 & 57.41 & 51.28 & 63.44 & 85.19 & 33.33 \\
85 & 58.70 & 70.97 & 70.37 & 71.79 & 65.59 & 62.96 & 69.23 & 63.44 & 74.07 & 48.72 \\
94 & 58.33 & 77.42 & 87.04 & 64.10 & 78.49 & 85.19 & 69.23 & 62.37 & 72.22 & 48.72 \\
102 & 57.61 & 57.61 & 56.60 & 58.97 & 60.87 & 66.04 & 53.85 & 59.78 & 71.70 & 43.59 \\
\hline
\textbf{Mean$\pm$SD} & \textbf{58.14$\pm$0.94} & \textbf{67.02$\pm$7.02} & 69.61$\pm$9.33 & 63.40$\pm$7.35 & \textbf{65.41$\pm$8.87} & 65.51$\pm$10.01 & 65.26$\pm$10.14 & \textbf{61.95$\pm$5.12} & 71.36$\pm$8.18 & 48.91$\pm$7.63 \\
\hline\\
\end{tabular}
}
\caption{Chance level accuracy compared to the performance of EEGNet model using 1) all 64 EEG channels, 2) using 21 Motor Imagery and movement-related EEG channels leveraging the domain knowledge, and 3) using 21 most relevant EEG channels identified by GradCAM.}
\label{tab:eegnet_accTable}
\end{table*}

\begin{figure*}[t!]
     \begin{subfigure}[]{0.32\textwidth}
         \centering
         \includegraphics[trim={15cm 0 12cm 2.2cm},clip,width=\linewidth]{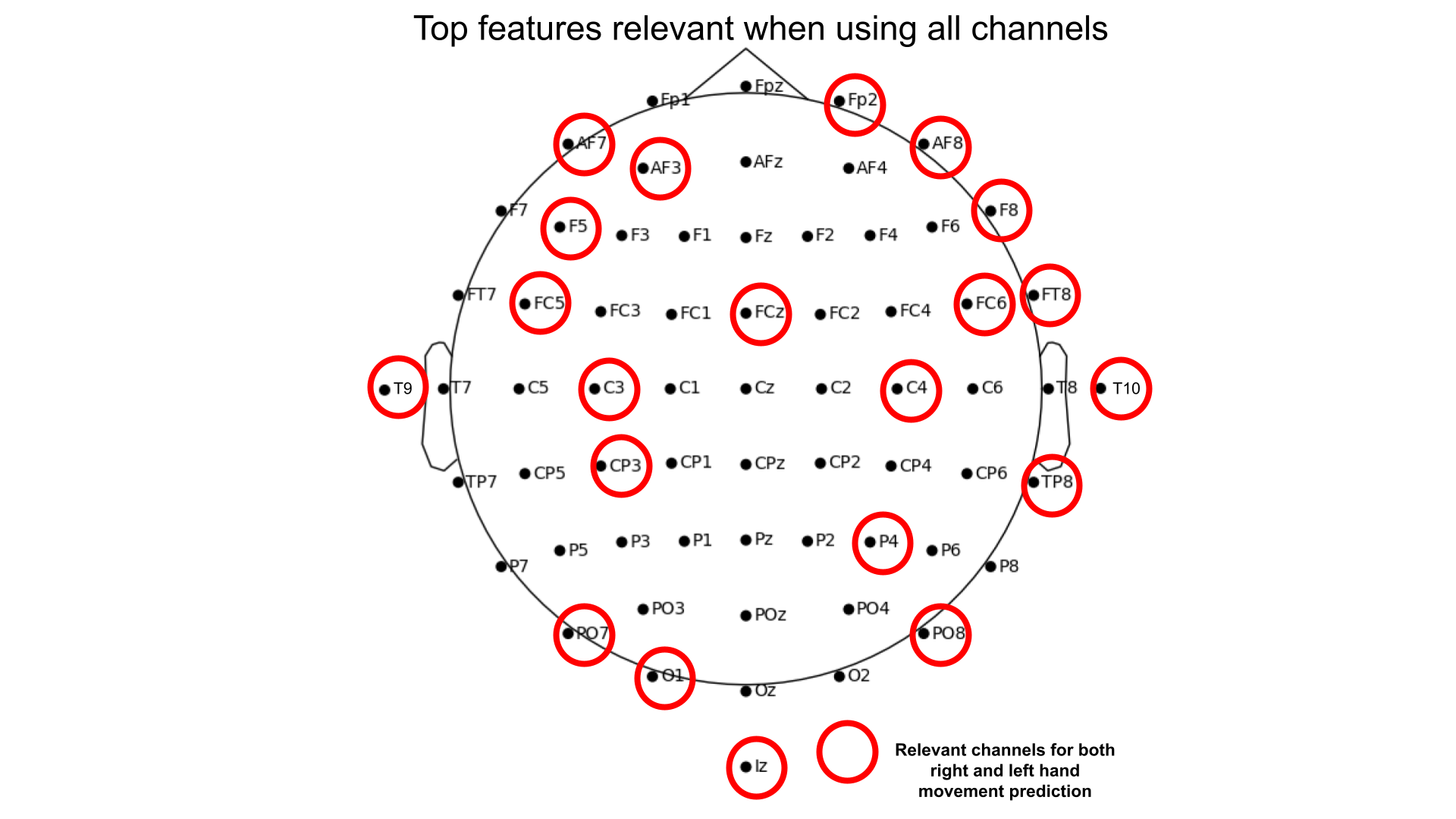}
         \caption{}
         \label{fig:riem_montage}
     \end{subfigure}
     \begin{subfigure}[]{0.32\textwidth}
         \centering
         \includegraphics[trim={15cm 0 12cm 2.2cm},clip,width=\linewidth]{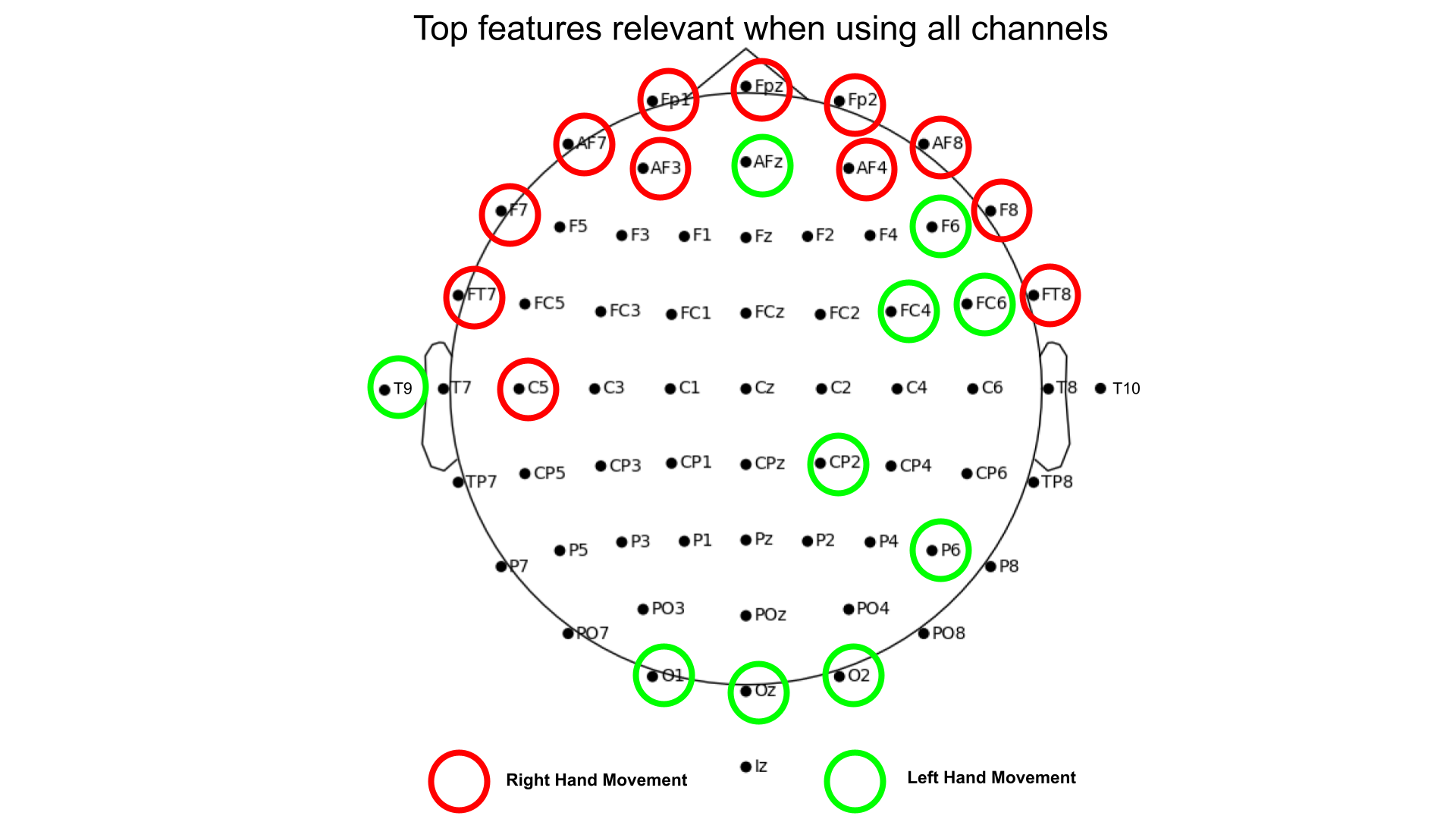}
         \caption{}
         \label{fig:conf_montage}
     \end{subfigure}
     \begin{subfigure}[]{0.32\textwidth}
         \centering
         \includegraphics[trim={15cm 0 12cm 2.2cm},clip,width=\linewidth]{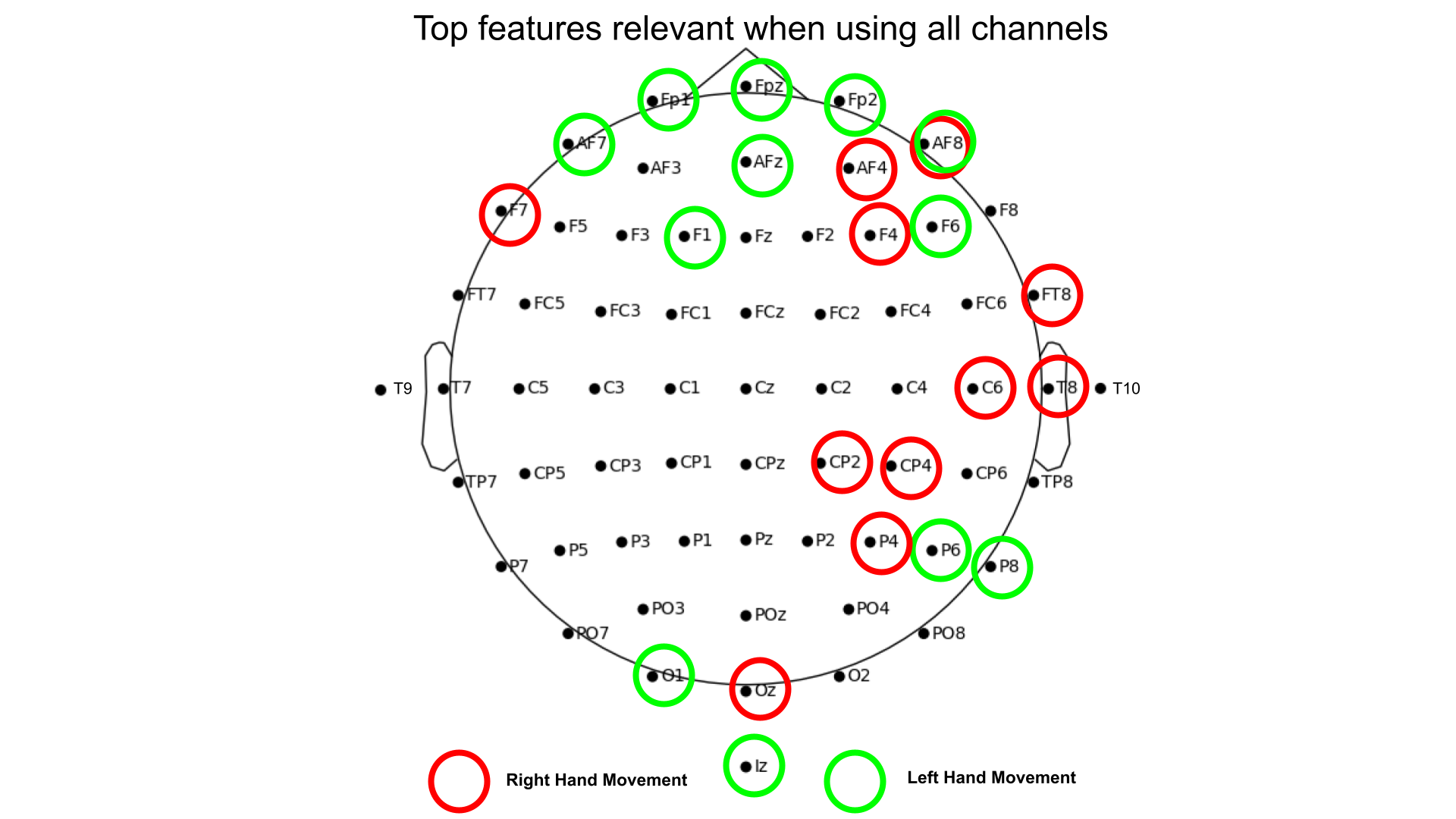}
         \caption{}
         \label{fig:eegnet_montage}
     \end{subfigure}
        \caption{Montage images highlighting the significant channels among the 64 channels, identified by a) using feature relevance scores based on Riemannian distance on covariance matrix, b) using GradCAM with EEGConformer model, and c) using GradCAM with EEGNet model.}
        \label{fig:montage_comparison}
\end{figure*}

\begin{figure*}[t!]
    \captionsetup[subfigure]{justification=centering}
     \begin{subfigure}[]{0.24\textwidth}
         \centering
         \includegraphics[trim={15cm 0 12cm 0cm},clip,width=\linewidth]{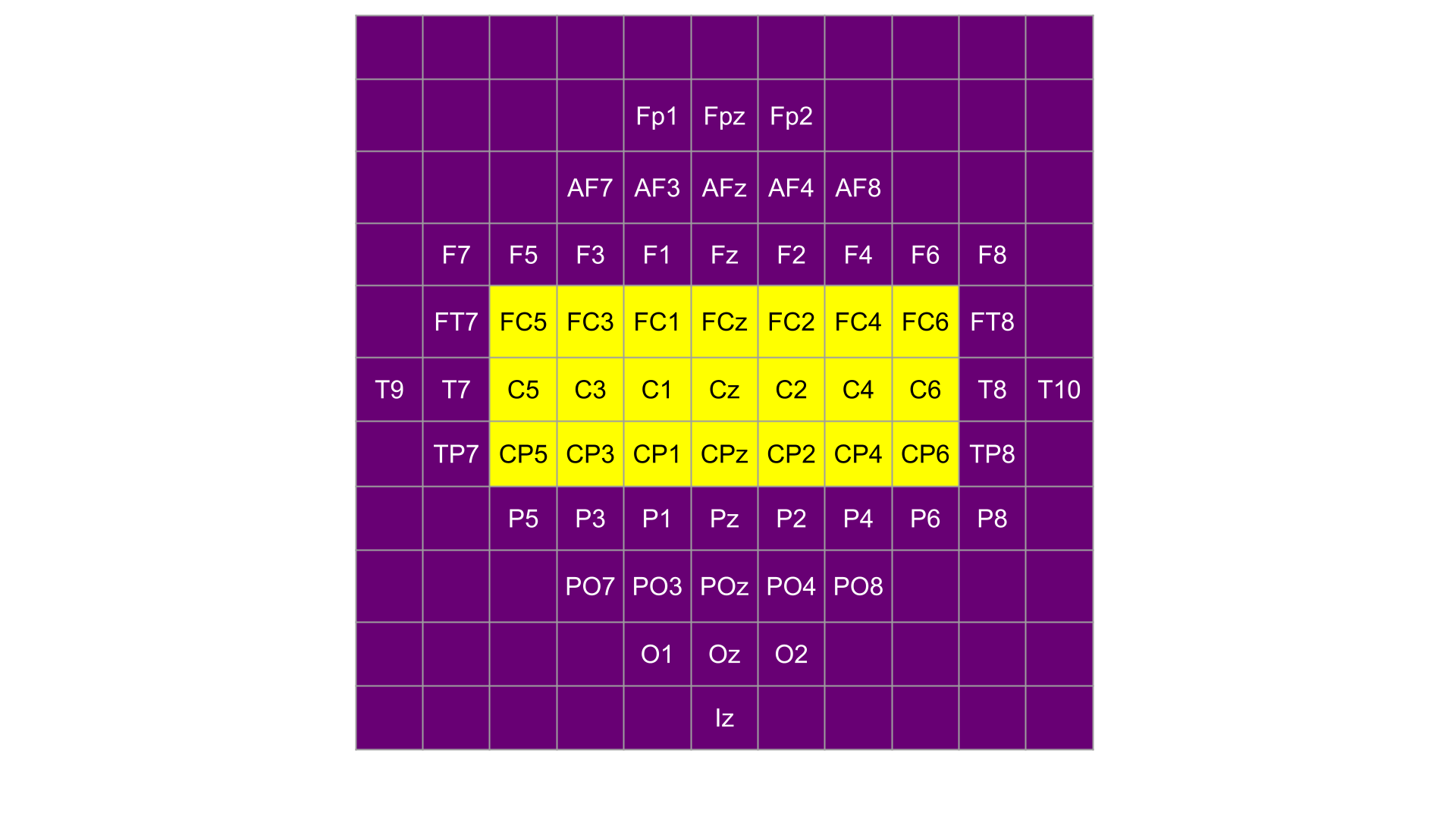}
         \caption{Baseline}
         \label{fig:mi_emd_map}
     \end{subfigure}
     \begin{subfigure}[]{0.24\textwidth}
         \centering
         \includegraphics[trim={15cm 0 12cm 0cm},clip,width=\linewidth]{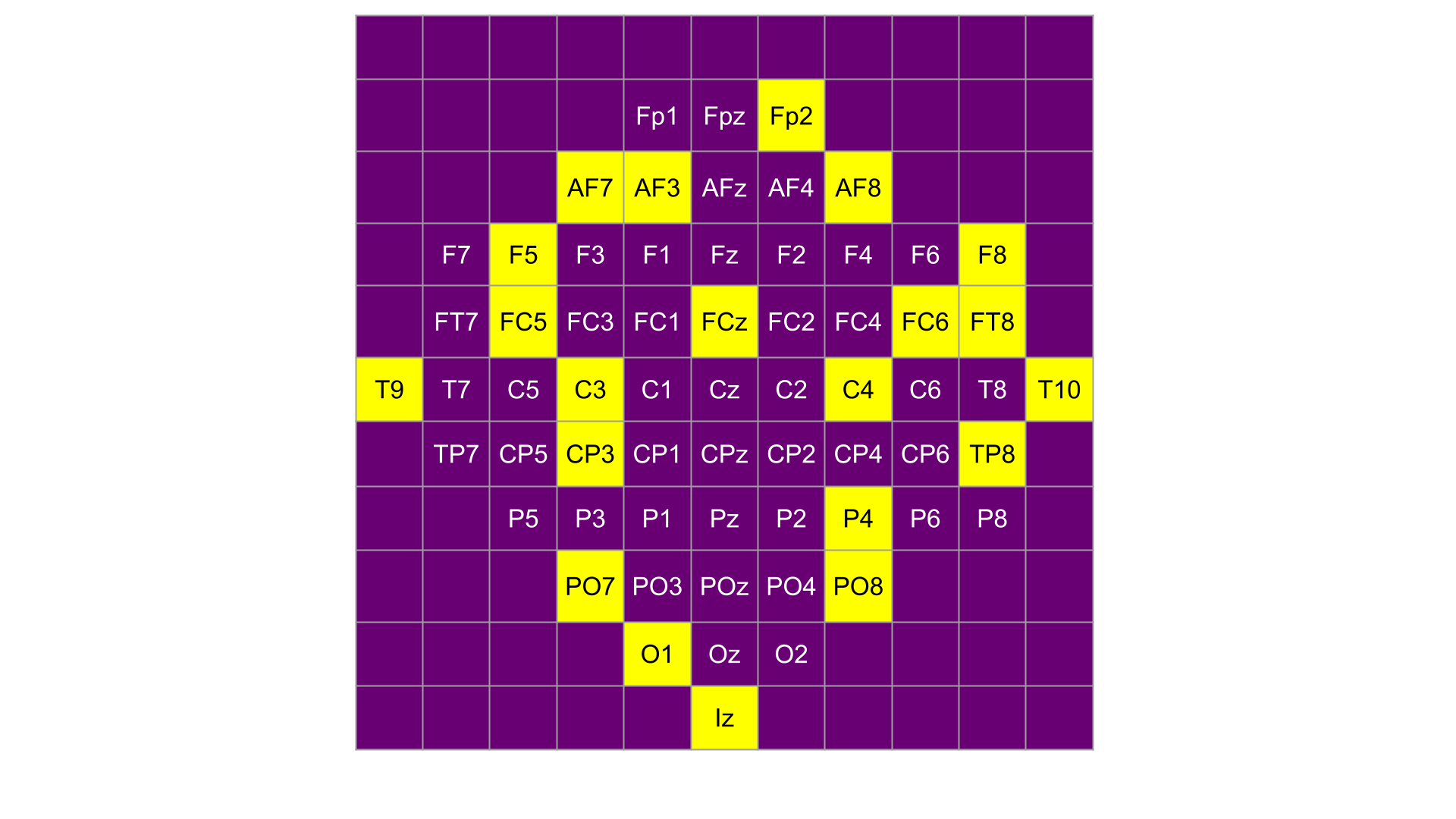}
         \caption{MDM-Riemannian \\ Earth mover's distance: 7.1612}
         \label{fig:riem_emd_map}
     \end{subfigure}
     \begin{subfigure}[]{0.24\textwidth}
         \centering
         \includegraphics[trim={15cm 0 12cm 0cm},clip,width=\linewidth]{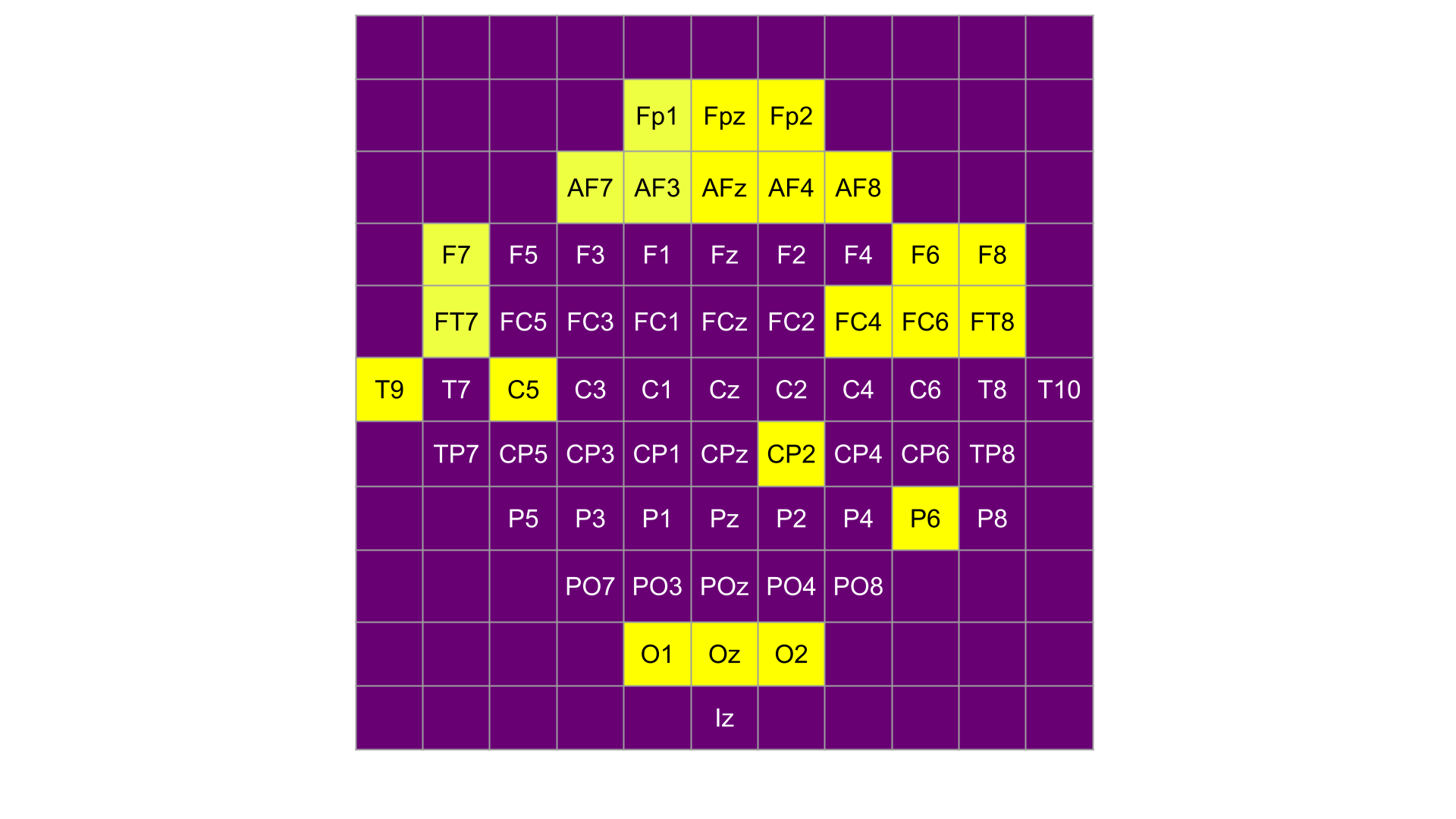}
         \caption{EEG Conformer \\ Earth mover's distance: 8.9289}
         \label{fig:conf_emd_map}
     \end{subfigure}
     \begin{subfigure}[]{0.24\textwidth}
         \centering
         \includegraphics[trim={15cm 0 12cm 0cm},clip,width=\linewidth]{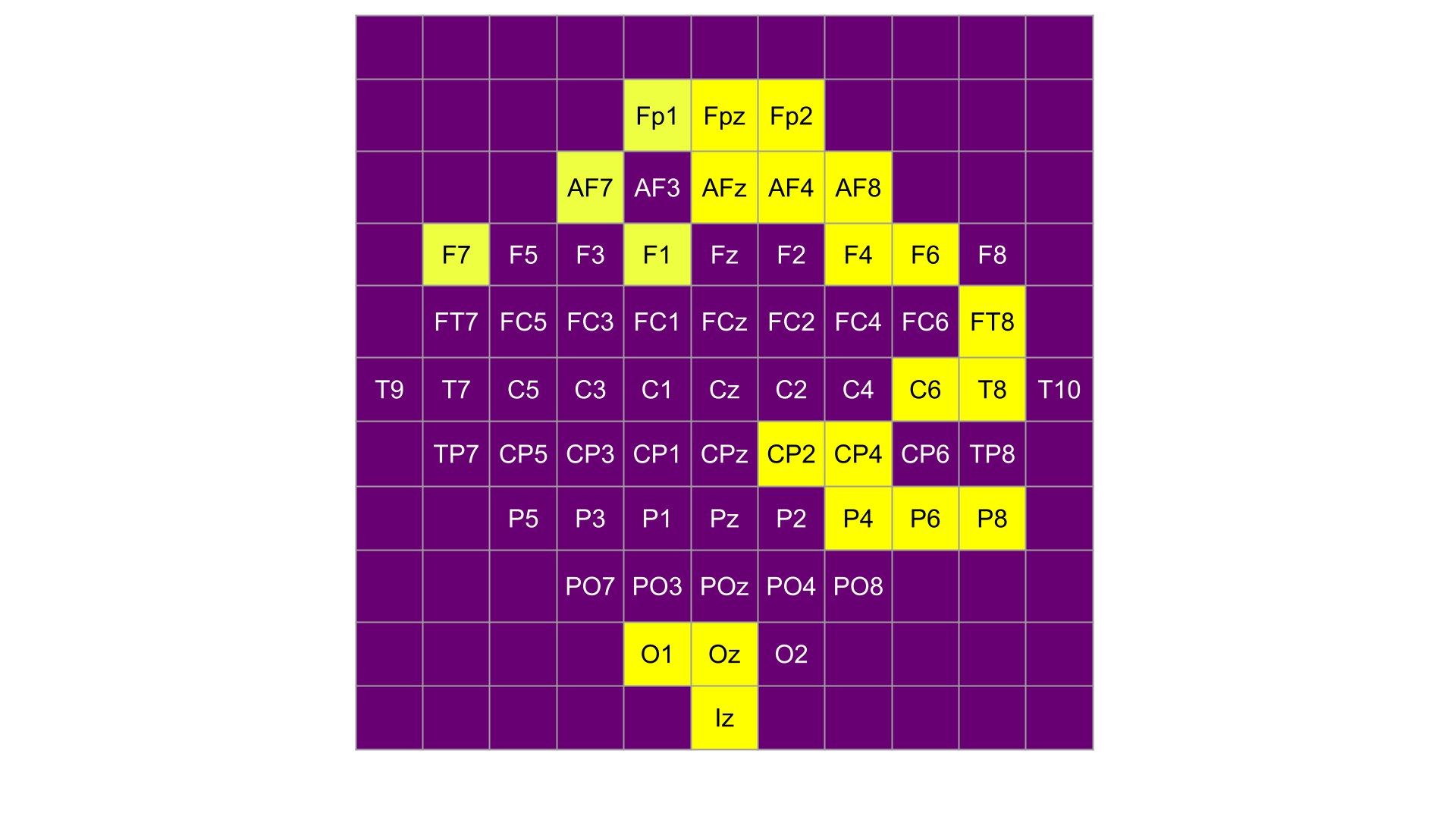}
         \caption{EEGNet\\ Earth mover's distance: 9.2948} 
         \label{fig:eegnet_emd_map}
     \end{subfigure}
        \caption{2D spatial representation of the significant channels among the 64 channels used for the comparison based on EMD. a) Baseline representation using 21 channels proximal to motor cortical regions, b) Relevance from MDM classifier using Riemannian distance is closest to baseline, b) followed by EEGConformer, and c) EEGNet model, when trained on top 14 participants. }
        \label{fig:spatial_map}
\end{figure*}

\begin{figure*}[t!]
    \captionsetup[subfigure]{justification=centering}
     \begin{subfigure}[]{0.23\textwidth}
         \centering
         \includegraphics[trim={15cm 0 12cm 0cm},clip,width=\linewidth]{mi_emd_map.png}
         \caption{Baseline}
         \label{fig:mi_cont_emd_map}
     \end{subfigure}
     \begin{subfigure}[]{0.24\textwidth}
         \centering
         \includegraphics[trim={15cm 0 10cm 0cm},clip,width=\linewidth]{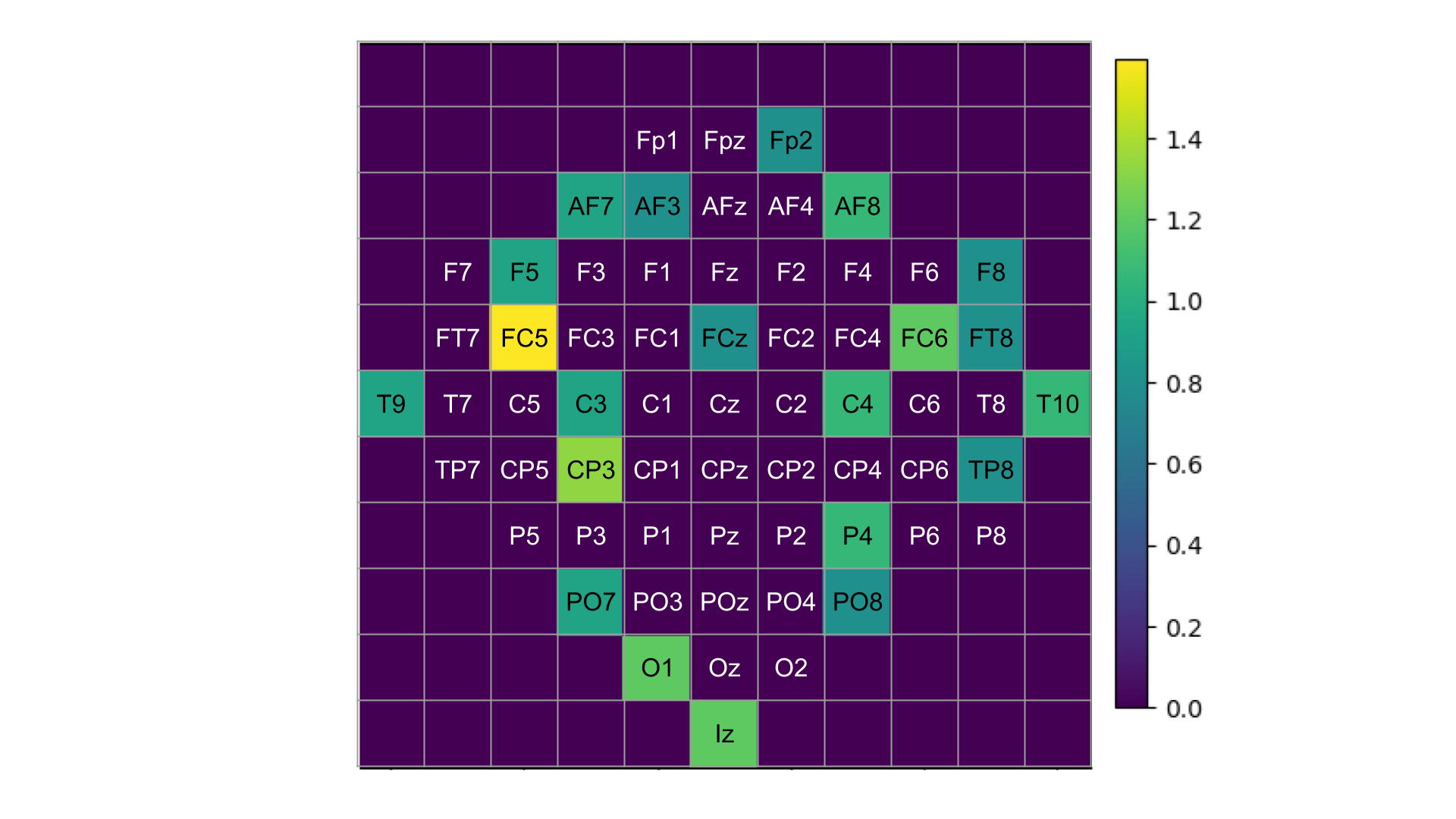}
         \caption{MDM-Riemannian \\ Earth mover's distance: 7.0931}
         \label{fig:riem_cont_emd_map}
     \end{subfigure}
     \begin{subfigure}[]{0.24\textwidth}
         \centering
         \includegraphics[trim={15cm 0 10cm 0cm},clip,width=\linewidth]{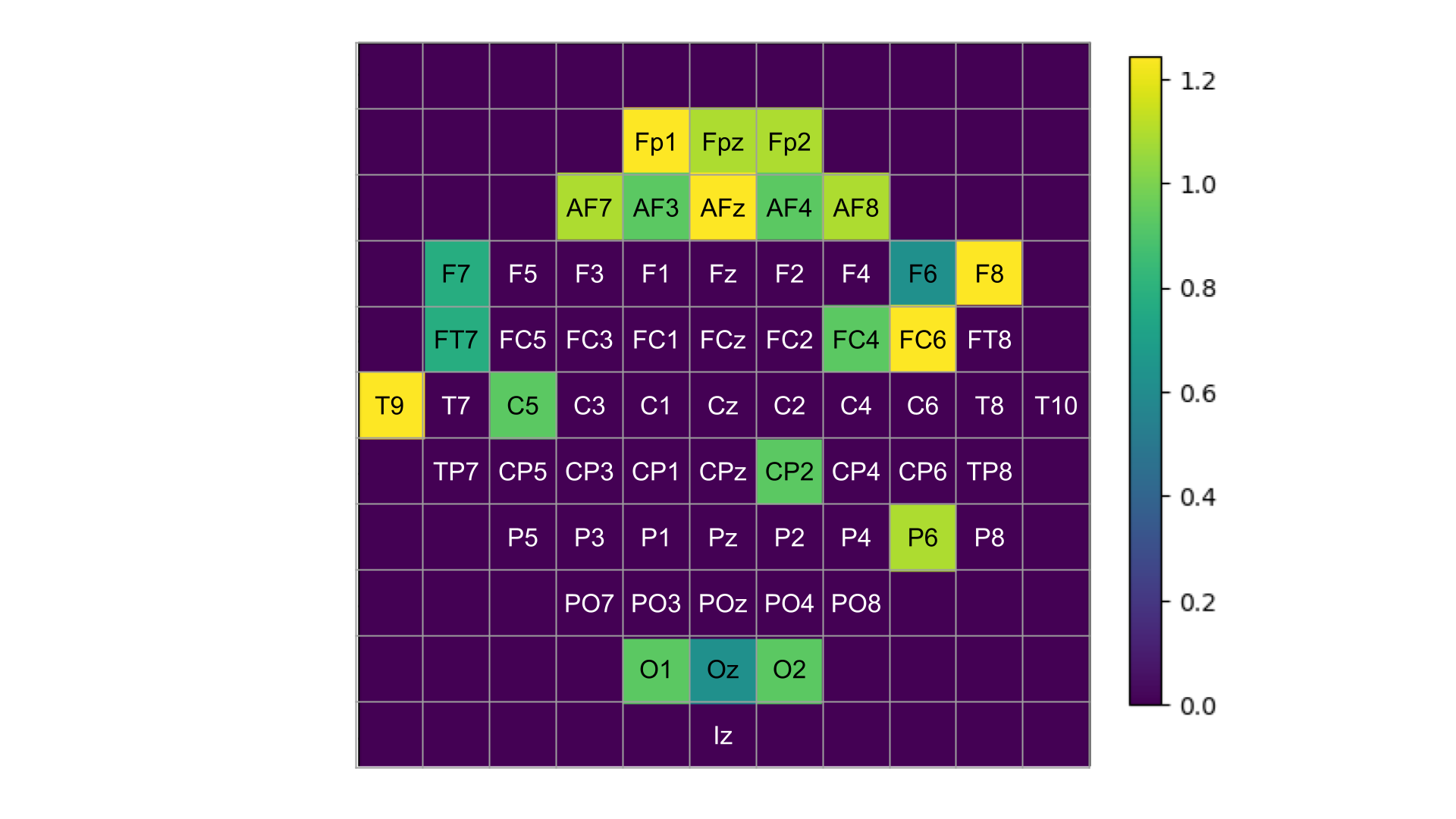}
         \caption{EEG Conformer \\ Earth mover's distance: 9.1471}
         \label{fig:conf_cont_emd_map}
     \end{subfigure}
     \begin{subfigure}[]{0.24\textwidth}
         \centering
         \includegraphics[trim={15cm 0 10cm 0cm},clip,width=\linewidth]{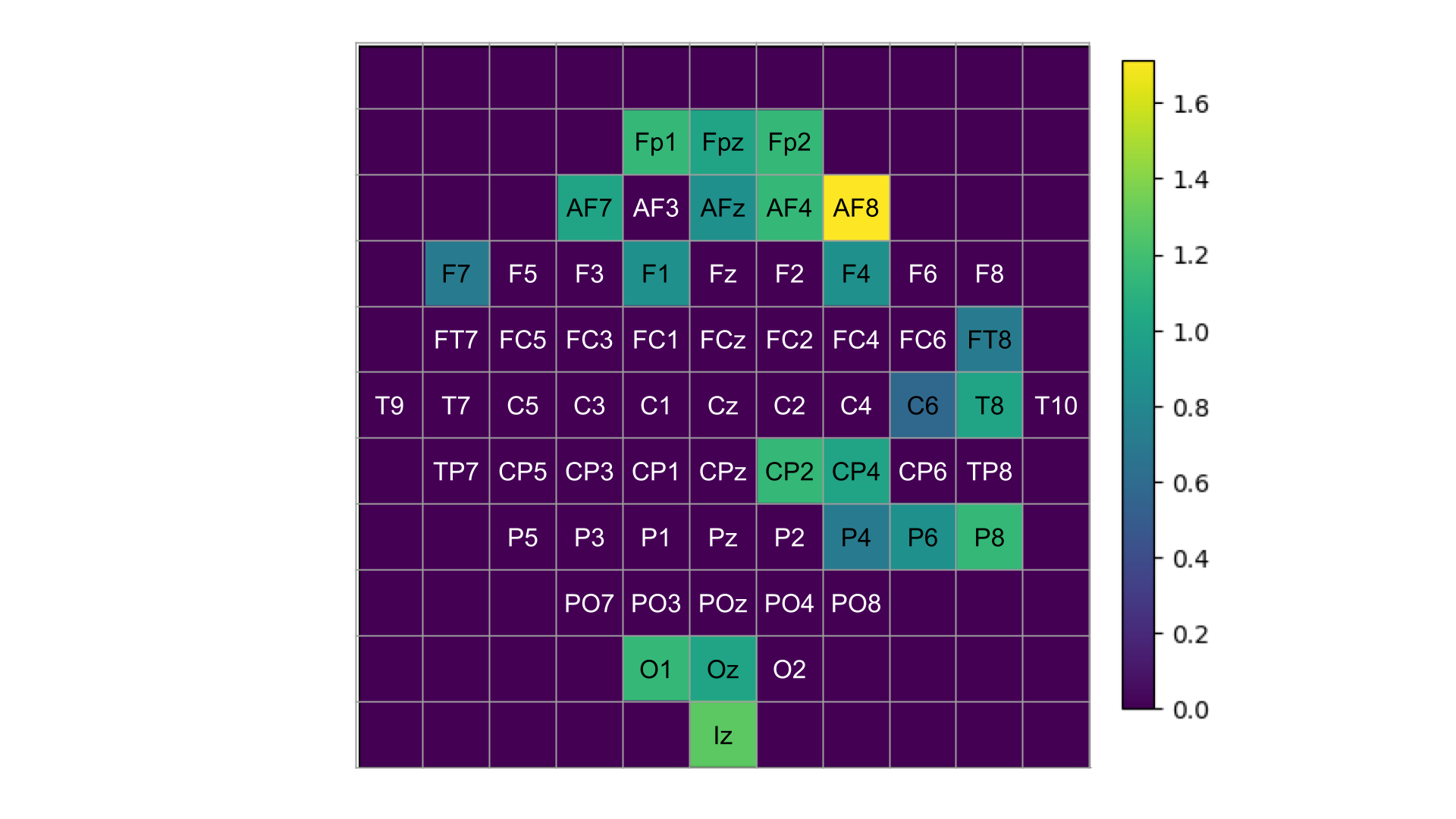}
         \caption{EEGNet\\ Earth mover's distance: 9.7759} 
         \label{fig:eegnet_cont_emd_map}
     \end{subfigure}
        \caption{2D spatial representation of the significant channels with weighted relevance among the 64 channels used for the comparison based on EMD. a) Baseline representation with equal importance to 21 channels proximal to motor cortical regions, b) Relevance from MDM classifier using Riemannian distance is closest to baseline, b) followed by EEGConformer, and c) EEGNet model, when trained on top 14 participants. }
        \label{fig:cont_spatial_map}
\end{figure*}

Comparisons of global-level model explanations are facilitated through montage images in \autoref{fig:montage_comparison}, allowing visualization of the distribution differences in feature importance across configurations. While channel importance for the MDM classifier is class agnostic, GradCAM provides channel importance for right and left-hand movement classification separately. 

The spatial maps represented in \autoref{fig:spatial_map} offer a concise representation of feature relevance and channel positioning, aiding in quantifying model explanations in the spatial domain. Following a similar trend in accuracy, the EMD for the MDM classifier is 7.1612, closest to the baseline condition of MI-relevant electrodes. Explanations from EEGConformer and EEGNet are at 8.9289 and 9.2948 distance, respectively, from the baseline. Considering the frequency of each channel in the top relevant features across participants, the weighted spatial map is represented in \autoref{fig:cont_spatial_map} for each model with distance from the baseline. Comparing the distances to unweighted spatial maps, EMD for the MDM classifier decreases to 7.0931 as it has greater weightage to channels in the subset of the baseline, while EEGConformer and EEGNet distances increase to 9.1471 and 9.7759, respectively.\\


\section{Discussion}
\label{sec:discussion}

The obtained results reveal intriguing insights into the performance of classification models and the feature relevance-driven explanations derived from channel selection or XAI techniques in the context of MI tasks.

The observed trend indicating Riemannian geometry outperforming EEGConformer and EEGNet showcases the effectiveness of Riemannian approaches in capturing the intrinsic geometry of symmetric positive definite (SPD) matrices, common in EEG covariance matrices. Features related to motor imagery are based on rhythms, not in the time domain, which the variance can also capture as it is proposed in the common spatial pattern (CSP)-based classifiers. However, since we chose the baseline model to be the MDM classifier, the results possibly ignored the participant's data where EEGConformer and EEGNet outperformed the Riemannian geometry-based approach.

The marginal performance advantage of EEGConformer over EEGNet is consistent with the nuanced differences in the neural network architecture. EEGConformer uses multi-head attention layers that help the network identify features across long sequences compared to EEGNet. The results motivate us to investigate feature relevance in other domains besides spatial information. Possible approaches could include comparisons in spectral and temporal domains by dividing each trial into shorter epochs and comparing feature relevance in those epochs.

The consistently better model performance of MI-relevant channels over feature-relevant channels, as demonstrated by GradCAM and backward elimination, adds depth to the understanding of spatial information. It underscores the significance of domain-specific knowledge in training data-driven models and carefully selecting channels that align with motor cortical regions.

The consistency of this trend in both accuracy and the calculated Wasserstein distance or EMD reinforces the reliability of the findings. The parallel outcomes across different metrics enhance the robustness of the conclusions. An interesting direction to explore is to understand neuroscience experts' opinions on reliability and their choice of model given the explanations. 


An interpretation of relevant channels that were significant in data-driven approaches may be linked to artefacts or activities that correlated with the task but were not causally involved. The fact that there are sensors that are needed and do not correspond to where the expected brain responses are projected on the scalp, holds for sensor-level analysis. However, an investigation to understand further the temporal and spectral domain explanations holds an interesting prospect to support such interpretations.



While the selection of MI-relevant channels based on the domain knowledge of specific regions associated with motor functions is a known practice, data-driven model outcomes need to be verified in the context where the signal-to-noise ratio is typically high and raw EEG data is used to train an end-to-end learning model. This insight has practical implications, especially in optimizing Brain-Computer Interface (BCI) systems. There is an increased interest in exploring neural network architectures optimising on Riemannian manifolds\cite{huang2017riemannian,kobler2022spd,pan2022matt,suh2021riemannian}. Therefore, an approach towards quantifying feature relevance and evaluation with respect to facts grounded in neuroscience is even more relevant.  


Explanations are used as tools for model improvement in BCIs\cite{rajpura2023explainable} and across domains \cite{weber2023beyond}. In this context, the proposed metric for comparing spatial maps can be employed in model regularisation, gradient, and feature augmentation. Concurrent efforts to understand the needs and the nuances of explanations for relevant stakeholders involved in the domain applications of BCIs are also required\cite{kim2024stakeholder}. Kim et al. \cite{kim2023designing} propose a theoretical framework for XAI interfaces that captures the requirements of a domain expert in BCI. Combining quantitative explanations with user-centric interfaces appears to be a promising approach towards reliable and trustworthy BCIs.

In the context of BCIs, where the interpretability of models is crucial for understanding and trusting the technology, the susceptibility to adversarial attacks could pose significant concerns. Galli et al.~\cite{galli2021reliability} highlight the susceptibility of DNNs and XAI outcomes to Adversarial Perturbations (APs), which are procedures designed to mislead a target model using imperceptible noise. Ravindran and Contreras-Vidal \cite{sujatha2023empirical} also conclude that XAI techniques also show reliability issues with simulated EEG data in the context of BCIs after randomizing either model weights or labels. Such works indicate the necessity to study the reliability of XAI outcomes vigorously. 

Acknowledging potential limitations, such as dataset-specific characteristics, is essential for this study. The study's generalizability will benefit from exploring diverse datasets to validate the observed trends across different experimental setups. Future work will include the analysis of other BCI paradigms, such as the event-related paradigm \cite{kubler2009brain,cecotti2015toward,lees2019speed}. It is indeed important to determine if a BCI is truly a BCI and not capturing muscle artefacts that are synchronized with the presentation of target stimuli. 

\section{Conclusion}
\label{sec:conclusion}

In this paper, we have analyzed and compared different state-of-the-art machine-learning algorithms for motor imagery classification. The results demonstrate that efficacy cannot be determined by using performance metrics like accuracy on limited datasets. The work proposed quantifying the explanations or feature relevance extracted using different techniques for three model architectures. The results confirm the performance of the Riemannian geometry-based classifier and suggest substantial discrepancies between the best sensors and where they are expected. The findings indicate that end-to-end data-driven approaches need to be validated in the context of BCIs, and the model interpretations should be compared against what can be expected from the literature by adding the XAI framework to the data-driven pipeline.

\section{Acknowledgment}
\label{sec:acknowledgment}
Our source code is publicly available on \href{https://github.com/HAIx-Lab/SpatialExplanations4BCI}{https://github.com/HAIx-Lab/SpatialExplanations4BCI}. This work was supported by Indian Institute of Technology Gandhinagar startup grant IP/IITGN/CSE/YM/2324/05. 

\bibliography{references}

\end{document}